\documentclass[10pt,romanappendices,conference,twocolumn]{IEEEtran}

\usepackage{times}
\usepackage{latexsym}
\usepackage{stfloats}
\usepackage{citesort}
\usepackage{graphicx, amsfonts,amsmath, amssymb,array,url,stfloats}
\usepackage[noadjust]{cite}
\usepackage{graphicx}
\usepackage{subfigure}
\usepackage{enumerate}

\newtheorem{corollary}{Corollary}
\newtheorem{proposition}{Proposition}
\newtheorem{lemma}{Lemma}

\begin{document}

\title{Impact of Residual Transmit RF Impairments on Training-Based MIMO Systems}
\author{\IEEEauthorblockN{Xinlin~Zhang\IEEEauthorrefmark{1}, Michail~Matthaiou\IEEEauthorrefmark{2}\IEEEauthorrefmark{1}, Mikael Coldrey\IEEEauthorrefmark{3},  and Emil~Bj\"ornson\IEEEauthorrefmark{4}\IEEEauthorrefmark{5}}
\IEEEauthorblockA{\IEEEauthorrefmark{1}Department of Signals and Systems, Chalmers University of Technology, Gothenburg, Sweden}
\IEEEauthorblockA{\IEEEauthorrefmark{2}School of Electronics, Electrical Engineering and Computer Science, {Queen's} University Belfast, Belfast, U.K.}
\IEEEauthorblockA{\IEEEauthorrefmark{3}Ericsson Research, Ericsson AB, Gothenburg, Sweden}
\IEEEauthorblockA{\IEEEauthorrefmark{4}Alcatel-Lucent Chair on Flexible Radio, SUPELEC, Gif-sur-Yvette, France}
\IEEEauthorblockA{\IEEEauthorrefmark{5}Department of Signal Processing, ACCESS, KTH Royal Institute of Technology, Stockholm, Sweden}

E-mail: xinlin@chalmers.se, m.matthaiou@qub.ac.uk, mikael.coldrey@ericsson.com, emil.bjornson@supelec.fr}
\maketitle

\begin{abstract}

Radio-frequency (RF) impairments, that exist intimately in wireless communications systems, can severely degrade the performance of traditional multiple-input multiple-output (MIMO) systems. Although compensation schemes can cancel out part of these RF impairments, there still remains a certain amount of impairments. These residual impairments have fundamental impact on the MIMO system performance. However, most of the previous works have neglected this factor. In this paper, a training-based MIMO system with residual transmit RF impairments (RTRI) is considered. In particular, we derive a new channel estimator for the proposed model, and find that RTRI can create an irreducible estimation error floor. Moreover, we show that, in the presence of RTRI, the optimal training sequence length can be larger than the number of transmit antennas, especially in the low and high signal-to-noise ratio (SNR) regimes. An increase in the proposed approximated achievable rate is also observed by adopting the optimal training sequence length. When the training and data symbol powers are required to be equal, we demonstrate that, at high SNRs, systems with RTRI demand more training, whereas at low SNRs, such demands are nearly the same for all practical levels of RTRI.

\end{abstract}
\thispagestyle{empty}

\section{Introduction}

MIMO point-to-point systems offer wireless communication with high data rates, without requiring additional bandwidth or transmit power. The pioneering works of \cite{telatar1999capacity} and \cite{foschini1998a} illustrated a linear growth in capacity in rich scattering environments by deploying more antennas at both the transmitter and receiver sides. However, to fully reap the advantages that MIMO systems can offer, instantaneous channel state information (CSI) is essential, especially at the receiver.

In practical systems, a training-based (or pilot-based) transmission scheme is usually utilized to estimate the channel and thereafter to transmit/receive data. This area is well covered in the literature (e.g., \cite{hassibi2003, tong2004pilot, biguesh2006training, mikael1, emil2010est, agarwal2012adaptive}); however, most of these works assume ideal RF hardware, which is quite unrealistic in practice. RF impairments, such as in-phase/quadrature-phase (I/Q) imbalance, high power amplifier non-linearities, and oscillator phase noise, are known to have a detrimental impact on practical MIMO systems \cite{schenk2008rf,studer2010residual}. Even though one can resort to calibration schemes to mitigate part of these impairments \cite{schenk2008rf}, there still remains a certain amount of residual distortions unaccounted for. These residual impairments stem from, for example, inaccurate models which are used to characterize the impairments, as well as, errors in the estimation of impairments' parameters. To the best of our knowledge, the only paper that considers training-based MIMO systems with residual impairments is \cite{emil2013est}. The authors therein analyzed the impact of impairments on the uplink channel estimation in a massive MIMO configuration. They reported an estimation error floor, and observed that by increasing the number of pilot symbols, one can average out the impact of impairments. However, they did not provide detailed power allocation and training sequence schemes, which are of pivotal importance in training-based point-to-point communication systems.

Motivated by the above discussion, we hereafter assess the impact of RTRI on training-based MIMO systems. More specifically, we first evaluate how RTRI affect channel estimation in the estimation phase, and observe an estimation error floor in the high SNR regime, which is analytically deduced. After that, we analyze an approximation for the achievable rate, using the classical technique of \cite{hassibi2003}, in the presence of channel estimation errors, as well as, residual distortions in the data transmission phase. Through optimizing power allocation and training sequence length, we find that, the optimal training duration can be larger than the number of transmit antennas, especially for low and high SNR values. Moreover, for more practical systems, which have the same transmit power per channel use during the estimation and data transmission phases, our results indicate that systems with higher RTRI require more training at high SNRs, whilst at low SNRs, the training demands almost the same for all practical levels of RTRI.

\textit{Notation:} Upper and lower case boldface letters denote matrices and vectors, respectively. The trace of a matrix is expressed by $\mathrm{tr}\left\{\cdot\right\}$. The $n \times n$ identity matrix is represented by $\mathbf I_n$. The expectation operation is $\mathbb {E} [\cdot]$, while the matrix determinant is denoted by det$(\cdot)$. The superscripts $(\cdot)^ H$ and $(\cdot)^{-1}$ stand for Hermitian transposition and matrix inverse, respectively. The Frobenius norm is denoted by $\left\|\cdot\right\|_F^2$. The symbol $\mathcal{CN}\left(\mathbf m, \boldsymbol\Sigma\right)$ denotes a circularly-symmetric complex multi-variate Gaussian distribution with mean $\mathbf m$ and covariance $\boldsymbol\Sigma$, while $\triangleq$ refers to ``is defined as''.

\section{Signal and system models}
In this paper, we consider a block fading channel with a coherence time of ${T}$ channel uses. During each block, the channel is constant, and is a realization of the uncorrelated Rayleigh fading model. Channel realizations between different blocks are assumed to be independent.

\subsection{System Model With Residual Transmit RF Impairments}

RF impairments  exist widely in practical wireless communication systems. Due to these impairments, the transmitted signal is distorted during the transmission processing,  hence cause a mismatch between the intended signal and what is actually transmitted. Even though compensation schemes are usually adopted to mitigate the effects of these impairments, there is always some amount of residual impairments. In \cite{schenk2008rf, studer2010residual}, the authors have shown that these residual impairments on the transmit side act as additive noise. Furthermore, experimental results in \cite{studer2010residual} revealed that such RTRI behave like zero-mean complex Gaussian noise, but with the important property that their average power is proportional to the average signal power. For sufficient decoupling between different RF chains, such impairments are statistically independent across the antennas. Moreover, impairments during different channel uses are also assumed to be independent. We now denote the RTRI noise as $\boldsymbol\Delta$. Then, the input-output relationship of a training-based MIMO system with $N_t$ transmit antennas and $N_r$ receive antennas within a block of $T$ symbols, can be expressed as
\begin{equation}
\mathbf Y = \sqrt{\frac{\rho}{N_t}}\mathbf H (\mathbf S + \boldsymbol \Delta) + \mathbf V, \label{eq:SysModelAll}
\end{equation}
where ${\mathbf S}\in\mathbb{C}^{N_t\times T}$ is the transmitted signal, $\rho$ is the average SNR at each receive antenna, and $\mathbf H\in\mathbb C^{N_r\times N_t}$ is the channel matrix. The receiver noise and the received signal are denoted as $\mathbf V\in\mathbb{C}^{N_r\times T}$ and $\mathbf Y\in\mathbb{C}^{N_r\times T}$, respectively. Each element of $\mathbf H$ and $\mathbf V$ follows an independent $\mathcal{CN}(0,1)$ distribution. We also assume that the entries of $\mathbf S$ have unit variance, so that $\rho$ is the average received SNR at each receive antenna. At last, according to the above discussion, we can characterize the RTRI noise $\boldsymbol\Delta\in\mathbb C^{N_t\times T}$ as
\begin{align}
\boldsymbol\Delta_{(i)} \sim \mathcal{CN}\left(\mathbf{0}, \delta^2 \mathbf I_{N_t}  \right), \mathbb E\left[\boldsymbol\Delta_{(i)}\boldsymbol\Delta_{(j)}^H\right] = \mathbf 0\notag\\
i, j= 1, 2, \dots, T, i\neq j,
\end{align}
where $\boldsymbol\Delta_{(i)}$ denotes the $i$-th column of $\boldsymbol\Delta$. The proportionality parameter $\delta$ characterizes the level of residual impairments in the transmitter. Note that $\delta$ appears in practical applications as the error vector magnitude (EVM) \cite{holma2011LTE}, which is commonly used to measure the quality of RF transceivers. For instance, 3GPP LTE has EVM requirements in the range $\left[0.08,0.175 \right]$ \cite{holma2011LTE}. The relationship between $\delta$ and EVM is defined as
\begin{equation}
\mathrm{EVM} \triangleq \sqrt{\frac{\mathbb{E}_{\boldsymbol\Delta}\left[ \left\| \boldsymbol\Delta \right\|_F^2 \right]}{\mathbb{E}_{\mathbf S}\left[ \left\| \mathbf S\right\|_F^2 \right]}} = \delta.
\end{equation}
When $\delta = 0$, it indicates ideal hardware implementation.

We can now decompose the system model in (\ref{eq:SysModelAll}) into training phase and data transmission phase as follows:
\subsubsection{Training Phase}

\begin{align}\label{eq:SysModelTraining}
\mathbf Y_p = \sqrt{\frac{\rho_p}{N_t}}\mathbf H \left(  \mathbf S_p + \boldsymbol\Delta_p \right) + \mathbf V_p, \mathrm{tr}\{\mathbf S_p^H\mathbf S_p\} = N_tT_p, 
\end{align}
where $\mathbf S_p\in\mathbb C^{N_t\times T_p}$ is the deterministic matrix of training sequences and is known by the receiver, $\rho_p$ is the average SNR during the training phase, and $\mathbf Y_d$ is the $N_r\times T_p$ received matrix. The distortion noise caused by the RTRI is characterized as
\begin{align}
{\boldsymbol\Delta_p}_{(i)}\sim \mathcal{CN}\left(\mathbf 0, \delta^2\mathbf I_{N_t}\right), \mathbb E\left[{\boldsymbol\Delta_p}_{(i)}{\boldsymbol\Delta_p}_{(j)}^H \right]= \mathbf 0,\notag \\i,j=1,2,\dots,T_p,i\neq j.
\end{align}
Note that this model is mathematically similar to the systems which use a superimposed pilot scheme \cite{mikael1}, where part of the data symbol is conveyed during the training phase, and acts like noise.

\subsubsection{Data Transmission Phase}
\begin{align}\label{eq:SysModelData}
\mathbf Y_d = \sqrt{\frac{\rho_d}{N_t}}\mathbf H \left( \mathbf S_d \!+ \!\boldsymbol\Delta_d \right) \!+\! \mathbf V_d, \mathbb{E}\Big[\mathrm{tr}\{\mathbf {S}_d^H\mathbf S_d\}\Big] = N_tT_d,
\end{align}
where $\mathbf S_d \in \mathbb{C}^{N_t \times T_d}$ is the matrix of data symbols with $\mathcal{CN}(0,1)$ entries,  $\rho_d$ is the average SNR during the data transmission phase, and $\mathbf Y_d$ is the $N_r \times T_d$ received signal matrix. The distortion noise caused by the RTRI during this phase is characterized as \begin{align}
{\boldsymbol\Delta_d}_{(i)}\sim \mathcal{CN}\left(\mathbf 0, \delta^2\mathbf I_{N_t}\right), \mathbb E\left[{\boldsymbol\Delta_d}_{(i)}{\boldsymbol\Delta_d}_{(j)}^H \right]= \mathbf 0,\notag \\i,j=1,2,\dots,T_d,i\neq j.
\end{align}
Recall that conservation of time and energy yields
\begin{equation}
T = T_p + T_d,~~ \rho T = \rho_p T_p + \rho_d T_d.
\end{equation}

The models in (\ref{eq:SysModelAll}), (\ref{eq:SysModelTraining}), and (\ref{eq:SysModelData}) include the characteristics of RTRI, and enable us to identify some fundamental differences in the training-based MIMO systems as compared to the ideal hardware case of \cite{hassibi2003}.

\section{LMMSE Channel Estimation}\label{sec:estimation}
In this section, we analyze the impact of RTRI on the channel estimation phase. Channel estimation is carried out during the first $T_p$ channel uses. Within each block, the estimator compares the received signal $\mathbf Y_p$ with the predefined training sequence matrix $\mathbf S_p$. The classical results on training-based channel estimation consider Rayleigh fading channels, which have independent complex Gaussian noise with known statistics \cite{hassibi2003, emil2010est}. However this is not the case herein since the distortion noise $\boldsymbol\Delta_p$ depends on the unknown random channel $\mathbf H$ through the multiplication $\mathbf H\boldsymbol\Delta_p$. Although the distortion noise is Gaussian when conditioned on a channel realization, the effective distortion is the product of Gaussian variables. Thus, it has a $\textit{complex double Gaussian distribution}$ \cite{don2012double}, which does not admit tractable manipulations.

We now derive the LMMSE estimator of $\mathbf H$ under the model in (\ref{eq:SysModelTraining}), which is given by the following lemma.
\begin{lemma}\label{theorem:estimator}
Given the received signal $\mathbf Y_p$ and the RTRI level $\delta$, the LMMSE estimator of $\mathbf H$  is
\begin{equation}\label{eq:estimator}
\hat{\mathbf H} = \mathbf Y_p\bigg( \mathbf S_p^H{\mathbf S_p} + \left( \delta^2\rho_p + 1 \right)\mathbf I_{T_p} \bigg)^{-1}\mathbf S_p^H.
\end{equation}
\end{lemma}
\begin{IEEEproof}
Since the rows of $\mathbf Y_p$ are independent and identically distributed (i.i.d.), we can write the LMMSE estimator in the general form $\hat{\mathbf{H}} = \mathbf Y_p \mathbf A$, where $\mathbf A$ should minimize the mean square error (MSE), which is defined as $\mathrm{MSE} \triangleq \mathrm{tr}\left(\mathbf C_e\right)$. Herein, $\mathbf C_e\!\triangleq\!\mathbb E\left[ \mathbf H_e^H\mathbf H_e\right]$ is defined as the estimation error covariance matrix, where $\mathbf H_e \triangleq \mathbf H - \hat{\mathbf H}$ is the estimation error matrix. The estimator in (\ref{eq:estimator}) is found by taking the first derivative of the MSE with respect to $\mathbf A$, and equating the result to zero.
\end{IEEEproof}

\begin{corollary}
The training sequence matrix $\mathbf S_p$ that minimizes the MSE should satisfy
\begin{equation}\label{eq:Sp}
\mathbf S_p\mathbf S_p^H = T_p\mathbf I_{N_t}
\end{equation}
and the corresponding MSE is given by
\begin{equation}\label{eq:MSE_expression}
\mathrm{MSE} = \frac{N_rN_t}{1 + g} ~\mathrm{with}~ g \triangleq \frac{\rho_p T_p}{N_t(\rho_p\delta^2+1)}.
\end{equation}
\end{corollary}
\begin{IEEEproof}
This corollary can be proved by applying the Lagrange multiplier method \cite{bertsekas1999nonlinear} on the MSE, subject to the power constraint $\mathrm{tr}\{\mathbf S_p^H\mathbf S_p\} = N_tT_p$. The resulting estimation error covariance matrix becomes
\begin{equation}
\mathbf C_e = \frac{N_r}{1+g}\mathbf I_{N_t}.
\end{equation}

Since $\mathbf H_e$ has zero mean, the variance of its entries can be expressed as $\sigma_{\mathbf H_e}^2 = \frac{1}{N_rN_t}\mathrm{tr}\{\mathbf C_e\} = \frac{1}{1+g}$, which is also defined as the normalized  MSE. By the orthogonality principle of LMMSE estimators \cite{kaybook1}, each element in $\hat{\mathbf H}$ has a variance of $\sigma_{\hat{\mathbf H}}^2 = 1 - \sigma_{\mathbf H_e}^2 = \frac{g}{1+g}$.
\end{IEEEproof}

\begin{figure}[t]
\begin{centering}
\includegraphics [keepaspectratio,width=\columnwidth]{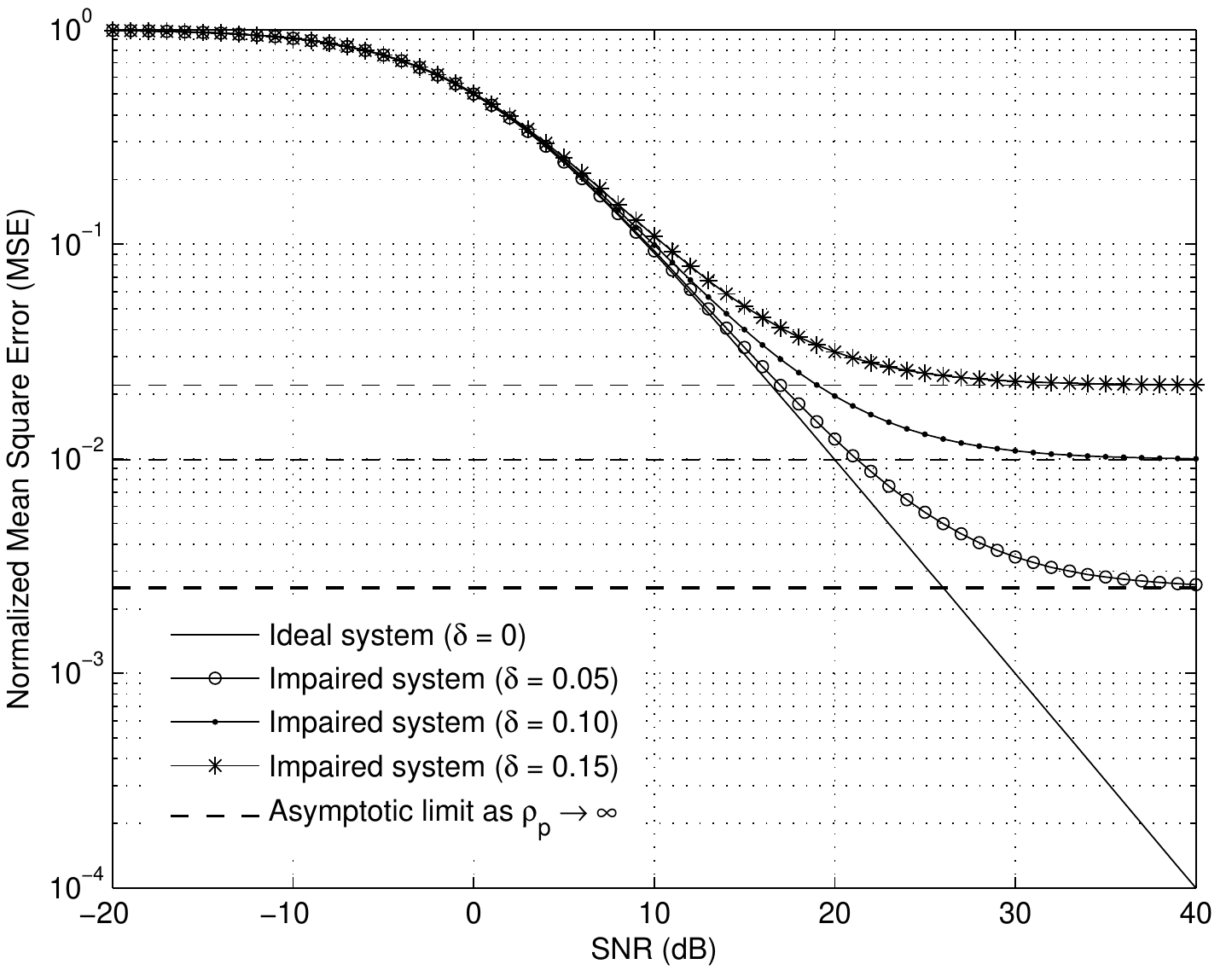} \vskip-2mm
\caption{Normalized mean square error (MSE) for different levels of impairments ($N_t = N_r = 4, T = 100, T_p = 4$).} \label{fig:MSE} \vskip-3mm
\end{centering}
\end{figure}

Figure \ref{fig:MSE} shows the normalized MSE, $\sigma_{\mathbf H_e}^2$, of a $4\times4$ MIMO system for different levels of impairments. In this case, we use $T_p\!=\!4$ channel uses to transmit pilot symbols, which is the minimum length required to estimate all channel dimensions. Without the existence of RTRI, increasing the transmit power decreases the MSE monotonically towards zero. However, in the presence of RTRI, we observe a fundamentally different behavior. Specifically, when the transmit power becomes high, impairments will generate an irreducible error floor, which is explicitly provided in the following corollary.
\begin{corollary}
Asymptotically as $\rho_p\!\rightarrow\!\infty$, the normalized MSE approaches the limit
\begin{equation}
\mathrm{MSE}_\mathrm{normalized}^{\rho_p\!\rightarrow\!\infty} = \frac{1}{1+\frac{T_p}{N_t\delta^2}}.\label{eq:MSE_limit}
\end{equation}
\end{corollary}
\begin{IEEEproof}
This corollary is simply achieved by making $\rho_p$ in (\ref{eq:MSE_expression}) large and normalize the MSE with respect to the number of transmit and receive antennas.
\end{IEEEproof}

Obviously, the value of this floor depends on the level of impairments; in general, large RTRI will cause severe degradation of the channel estimates. We can also see from (\ref{eq:MSE_limit}) that, for a fixed level of RTRI, an increase in the training sequence length $T_p$ decreases the MSE monotonically. As expected, for low SNR values, impairments have only limited impact, which is in line with the results of \cite{emil2013est}.

\section{Data Transmission}
This section analyzes the achievable rate of the non-ideal training-based MIMO system. The results in \cite{hassibi2003}, under the assumption of ideal hardware, are frequently used as reference.

During the data transmission phase, the estimated channel $\hat{\mathbf H}$ is available at the receiver. The receiver uses $\hat{\mathbf H}$ as if it were the true channel realization to recover the intended signal $\mathbf S_d$. Recalling that $\mathbf H\!=\!\hat{\mathbf H}\!+\! \mathbf H_e$, we may rewrite the received signal as
\begin{align}
\mathbf Y_d &= \sqrt{\frac{\rho_d}{N_t}}\hat{\mathbf H}\left(  \mathbf S_d + \boldsymbol\Delta_d \right) + \sqrt{\frac{\rho_d}{N_t}}\mathbf H_e\left(  \mathbf S_d + \boldsymbol\Delta_d \right) + \mathbf V_d\notag\\
&= \sqrt{\frac{\rho_d}{N_t}}\hat{\mathbf H}\mathbf S_d \!+\! \underbrace{\sqrt{\frac{\rho_d}{N_t}}\left(\hat{\mathbf H}\boldsymbol\Delta_d \!+\! \mathbf H_e\mathbf S_d \!+\! \mathbf H_e\boldsymbol\Delta_d\right)\! +\! \mathbf V_d}_{\tilde{\mathbf V}},
\end{align}
where $\tilde{\mathbf V}$ is the ``effective noise'' matrix. Note that each entry of $\tilde{\mathbf V}$ has zero-mean and the variance
\begin{align}
\sigma_{\tilde{\mathbf V}}^2 = \frac{1}{N_rT_d}\mathbb E\left[ \mathrm{tr}\left\{ {\tilde{\mathbf V}^H} \tilde{\mathbf V}  \right\}\right]
=\left(\frac{1}{1+g} + \delta^2\right)\rho_d + 1.
\end{align}
On a similar note, we can define $\bar{\mathbf H}\!\!\triangleq\!\!\tiny{\frac{1}{\sigma_{\hat{\mathbf H}}^2}}\!\hat{\mathbf H}$, which has uncorrelated and approximately $\mathcal{CN}(0,1)$ entries. \footnote{As we have emphasized in Section \ref{sec:estimation}, $\hat{\mathbf H}$ contains the multiplicative term $\mathbf H\boldsymbol\Delta_p$, which is complex double Gaussian distributed. This additional distortion, however, is insignificant for practical levels of RTRI; thus, the assumption of Gaussian distribution on the elements of $\hat{\mathbf H}$ is rather realistic.}

Given that $\hat{\mathbf H}$ is known to the receiver, it is straightforward to prove that $\mathbf S_d$ and $\tilde{\mathbf V}$ are uncorrelated. From \cite{hassibi2003}, we know that the worst-case effective noise is circularly-symmetric complex Gaussian distributed, with the same covariance as $\tilde{\mathbf V}$, Then, we can straightforwardly obtain a capacity lower bound as in \cite[Theorem 1]{hassibi2003}.

In the considered case though, where the channel estimate (\ref{eq:estimator}) contains the multiplicative term $\mathbf H\boldsymbol\Delta_p$, $\hat{\mathbf H}$ is only approximately Gaussian. Then, we can work out the approximated achievable rate according to
\begin{equation}
\tilde{R} = \frac{T_d}{T}\mathbb E\left[\mathrm{log_2}\ \mathrm{det}\!\left( \mathbf I_{N_r}+\rho_{\mathrm{eff}}\frac{\bar{\mathbf H}\bar{\mathbf H}^H}{N_t}\right) \right]\label{eq:CapacityLB},
\end{equation}
where  $\rho_{\mathrm{eff}}$ denotes the effective SNR,
\begin{align}
\rho_{\mathrm{eff}}&\triangleq \frac{\rho_d\sigma_{\hat{\mathbf H}}^2}{\sigma_{\tilde{\mathbf V}}^2}\\
&=\frac{\rho_d\rho_pT_p}{N_t(1+\rho_p\delta^2)(1+\rho_d+\rho_d\delta^2)+\rho_pT_p+\rho_d\rho_pT_p\delta^2}.\label{eq:effSNR}
\end{align}

\subsection{Optimizing over Power Allocation}
First, we optimize the power allocation to maximize the effective SNR $\rho_{\mathrm{eff}}$.

Let $\alpha$ denote the fraction of the total transmit power that is assigned to the data transmission phase. Then, we have
\begin{equation}
\rho_dT_d = \alpha\rho T, ~\rho_pT_p = (1-\alpha)\rho T, ~0 < \alpha < 1.
\end{equation}
\begin{proposition}
The optimal power allocation $\alpha\!\triangleq\!\frac{\rho_dT_d}{\rho T}$ in a training-based MIMO system with RTRI is given by
\begin{equation}\label{eq:ALPHA}
\alpha^{\mathrm{opt}} = \begin{cases}
\frac{r-\sqrt{r^2-rs}}{s}, & \textrm{for $s \neq 0$}\\
\frac{1}{2}, & \textrm{for $s = 0$}
\end{cases}
\end{equation}
where for concision, we have defined
\begin{align}
r&\triangleq  \rho T + \frac{N_t\rho T\delta^2}{T_p} + N_t,\notag\\
s&\triangleq  \rho T + \frac{N_t\rho T\delta^2}{T_p} - \frac{N_t\rho T(1+\delta^2)}{T_d}.\notag
\end{align}
\begin{IEEEproof}
Substituting $\rho_p\!=\!\frac{(1-\alpha)\rho T}{T_p}$ and $\rho_d\!=\!\frac{\alpha\rho T}{T_d}$ into (\ref{eq:effSNR}), then taking the first and second derivatives of $\rho_{\mathrm{eff}}$ with respect to $\alpha$ and equating the result to be zero, the proof follows immediately.
\end{IEEEproof}
\end{proposition}


Specifically, for high and low SNRs, we have
\begin{corollary}
At high and low SNRs, the optimal power allocation $\alpha$ reduces to
\begin{itemize}
\item At high SNRs, as $\rho\rightarrow\infty$
\begin{equation}\label{eq:alfahigh}
\alpha^{\mathrm{opt}} = \frac{\left(1+\frac{N_t\delta^2}{T_p}\right)\left(1+\sqrt{\frac{N_t(1+\delta^2)}{T_d}}\right)}{1+\frac{N_t\delta^2}{T_p}-\frac{N_t(1+\delta^2)}{T_d}},
\end{equation}
\item At low SNRs, as $\rho\rightarrow 0$
\begin{equation}\label{eq:alfalow}
\alpha^{\mathrm{opt}} = \frac{1}{2}.
\end{equation}
\end{itemize}
\end{corollary}

Clearly, at low SNR, half of the transmit power should be assigned to the training phase, which is consistent with the results of \cite{hassibi2003}. With the help of (\ref{eq:ALPHA}), we can further optimize the training length to maximize the approximated achievable rate.

\subsection{Optimizing over $T_p$}
In this part, we seek to determine the optimal training length $T_p$. Recall from \cite{hassibi2003} that, for ideal hardware systems over i.i.d. Rayleigh fading channels, it is always optimal to use as few channel uses as possible (i.e., $N_t$) for pilot symbols, regardless of the values of $\rho$ and $T$. However, for non-ideal hardware systems, we will show that this is no longer the case, since the optimal training length could be larger than $N_t$.

The standard way of finding the optimal training sequence length $T_p$ requires to substitute the optimal power allocation scheme $\alpha^{\mathrm{opt}}$ back to the approximated achievable rate in (\ref{eq:CapacityLB}), and then take the derivative of $\tilde{R}$ with respect to $T_p$. Unfortunately, this is not analytically tractable. To overcome this problem, we first derive the approximated achievable rate in (\ref{eq:CapacityLB}) in closed-form, which only depends on the values of SNR and $T_p$ for a given system setup ($N_t$, $N_r$, and $T$). Then, for each value of SNR, we can perform an exhaustive search over the integer $T_p$ to find the global optimum.

To facilitate our analysis, we herein present the following proposition.
\begin{proposition}
The approximated achievable rate in (\ref{eq:CapacityLB}), is analytically given by
\begin{equation}
\begin{split}
\tilde{R} &= \frac{qKT_d}{{\ln (2)T}}\sum\limits_{n = 1}^q {\sum\limits_{m = 1}^q {{{\left( { - 1} \right)}^{n + m}} \det \left( \boldsymbol\Omega  \right)} }  \Gamma \left(t\right)e^{\frac{N_t}{\rho_{\mathrm{eff}}}} \\
&\times\sum\limits_{k = 1}^{t}{\frac{\Gamma\left(-t+k,\frac{N_t}{\rho_{\mathrm{eff}}} \right)}{{\left(\frac{\rho_{\mathrm{eff}}}{N_t}\right)}^{t-k}}}\label{eq:analcapacity}
\end{split}
\end{equation}
where $q \triangleq \min(N_r,N_t)$, $p\triangleq \max(N_r,N_t)$ and $t \triangleq n+m+p-q-1$. Also, ${K} = \left[ \prod_{i=1}^{q}(p-i)!\prod_{j=1}^{q}(q-j)!  \right]^{-1}$ is a normalization constant. Moreover, $\Gamma(x)$ and $\Gamma(y,z)$  denote the Gamma function \cite[Eq.~(8.310.1)]{gradshteyn2007a} and the upper incomplete Gamma function \cite[Eq.~(8.350.2)]{gradshteyn2007a}, respectively. Finally, $\boldsymbol\Omega$ is a $(q-1)\times(q-1)$ matrix whose $(i,j)$-th element is given by
\begin{equation}
\boldsymbol\Omega_{i,j}=\left(\gamma_{i,j}^{(n)(m)} + p - q\right)! \, q^{-\frac{1}{q-1}}\notag
\end{equation}
where
\begin{equation}
\gamma_{i,j}^{(n)(m)} \triangleq \begin{cases}
i+j-2, & \textrm{if $i<n$ and $j<m$}\\
i+j, & \textrm{if $i\ge n$ and $j \ge m$}\\
i+j-1, & \textrm{otherwise.}
\end{cases}
\end{equation}

\begin{IEEEproof}
We can rewrite (\ref{eq:CapacityLB}) as
\begin{equation}
\tilde{R} = \frac{T_d}{T}\mathbb E\left[\mathrm{log_2det}\left( \mathbf I_{q}+\frac{\rho_{\mathrm{eff}}}{N_t}\mathbf W\right) \right],\label{eq:CapacityLB2}
\end{equation}
where $\mathbf W$ is defined as
\begin{equation} \label{eq:W-definition}
\mathbf W \triangleq \begin{cases}
\bar{\mathbf H}\bar{\mathbf H}^H, & \textrm{if $N_r\leq N_t$},\\
\bar{\mathbf H}^H\bar{\mathbf H}, & \textrm{if $N_r > N_t$}.
\end{cases}
\end{equation}
Note that $\mathbf W$ is a $q\times q$ random, non-negative definite matrix following the complex Wishart distribution. Thus, it has real non-negative eigenvalues and the probability density function (PDF) of its unordered eigenvalue, $\lambda$, is found in \cite[Eq.~(38)]{zanella2009marginal} to be
\begin{equation}
p_\lambda(\lambda) = K\sum\limits_{n = 1}^{q}\sum\limits_{m = 1}^{q} \frac{(-1)^{m+n} \lambda^{n+m+p-q-2}}{ e^{\lambda}} \det \left({\boldsymbol\Omega}\right).
\end{equation}

By exploiting the eigenvalue properties, we can now alternatively express the approximated achievable rate in (\ref{eq:CapacityLB2}) as
\begin{equation}\label{eq:int}
\tilde{R} =\frac{qT_d}{T}\int\limits_0^\infty\log_2\left(1+\frac{\rho_{\mathrm{eff}}}{N_t}\lambda\right)p_{\lambda}(\lambda)d\lambda.
\end{equation}
This integral can be evaluated using the integral identity in \cite[Eq.~(40)]{kang2006rician}.
The expression in (\ref{eq:analcapacity}) then follows after some simple algebraic manipulations.
\end{IEEEproof}
\end{proposition}

\begin{figure}[Ht!]
\begin{centering}
\includegraphics [keepaspectratio,width=\columnwidth]{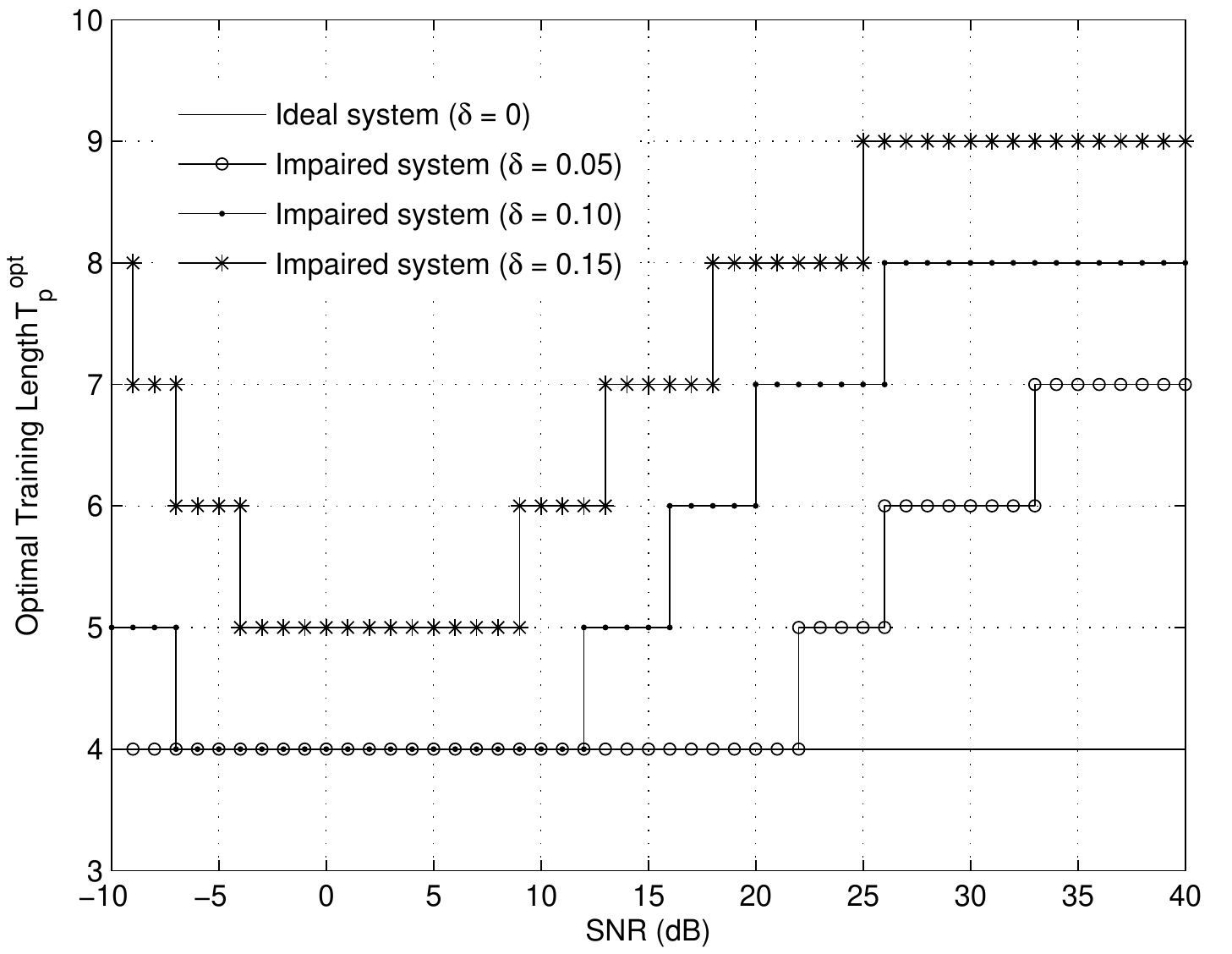} \vskip-2mm
\caption{Optimal training sequence length for different levels of impairments when both power allocation and training sequence length are optimized ($N_t = N_r = 4, T = 100$).} \label{fig:Tp_ALFA} \vskip-3mm
\end{centering}
\end{figure}

Based on (\ref{eq:analcapacity}), we perform an exhaustive search over the integer $T_p$ for different SNR values. Figure \ref{fig:Tp_ALFA} compares the optimal training sequence length, $T_p^{\mathrm{opt}}$, for the ideal and impaired systems. For the ideal hardware system, the optimal training length is always equal to the number of transmit antennas, which has already been proved in \cite{hassibi2003}. For the non-ideal hardware systems with RTRI, however, the optimal training sequence length may become larger than $N_t$. Generally speaking, higher impairment levels impose longer training sequences. At high SNRs, the effective SNR saturates, thus the overall performance cannot be improved by increasing the power; however, we can benefit by extending the training period. This is because the total pilot power is spread over $T_p$ channel uses, hence the impact of the temporally uncorrelated RTRI will be averaged over $T_p$ as well. It is also worth mentioning that in the low SNR regime, where thermal noise dominates the system performance, there is still an increase in achievable rate by improving the channel estimation with longer training sequences. The above results are valid for different number of antennas, and can be extended to massive MIMO systems with large receive antenna arrays.

\begin{figure}[Ht!]
\begin{centering}
\includegraphics [keepaspectratio,width=\columnwidth]{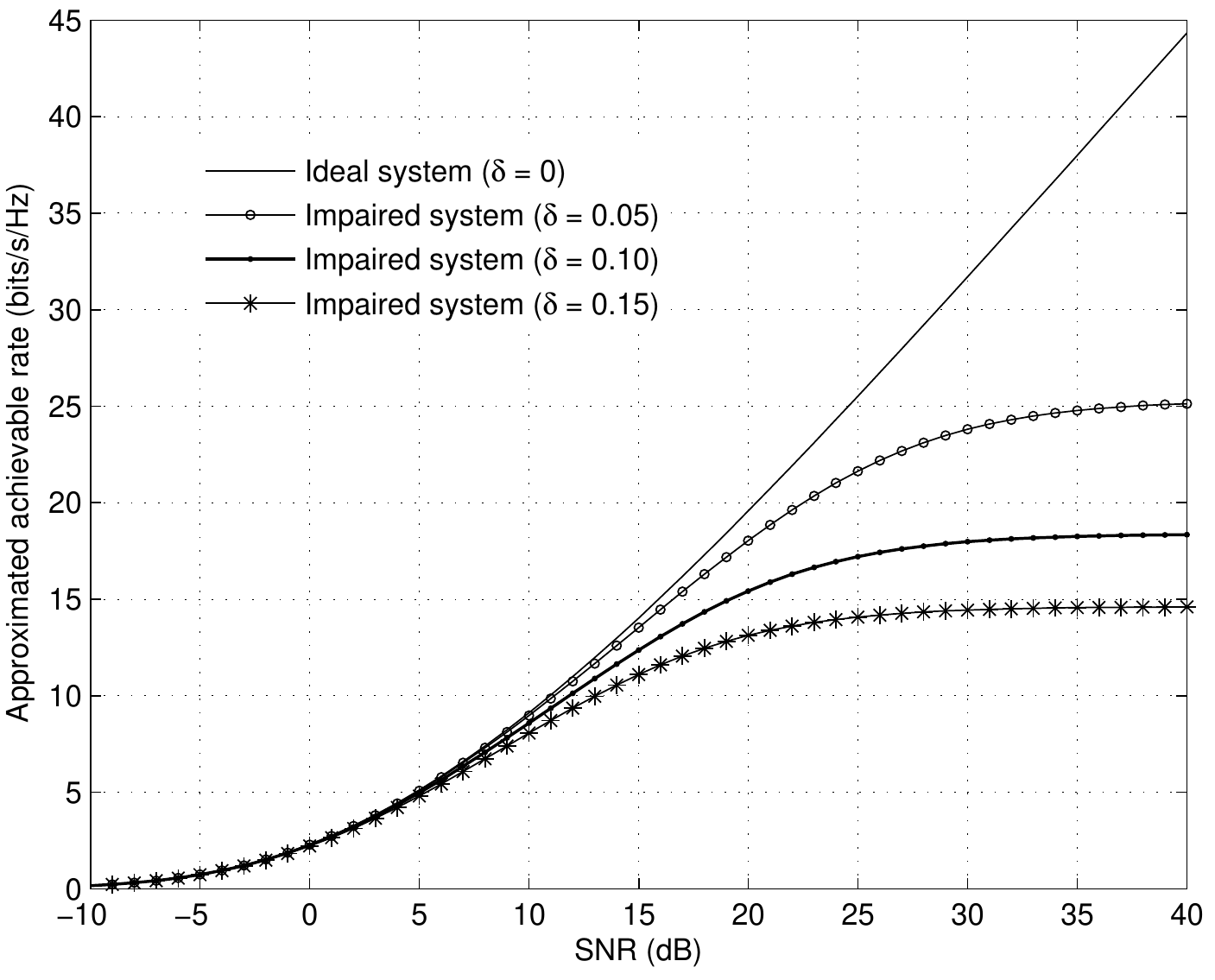} \vskip-2mm
\caption{The approximated achievable rate when both power allocation and training sequence length are optimized ($N_t = N_r = 4, T = 100$).} \label{fig:CAP_ALFA} \vskip-5mm
\end{centering}
\end{figure}

\begin{figure}[Ht!]
\begin{centering}
\includegraphics [keepaspectratio,width=\columnwidth]{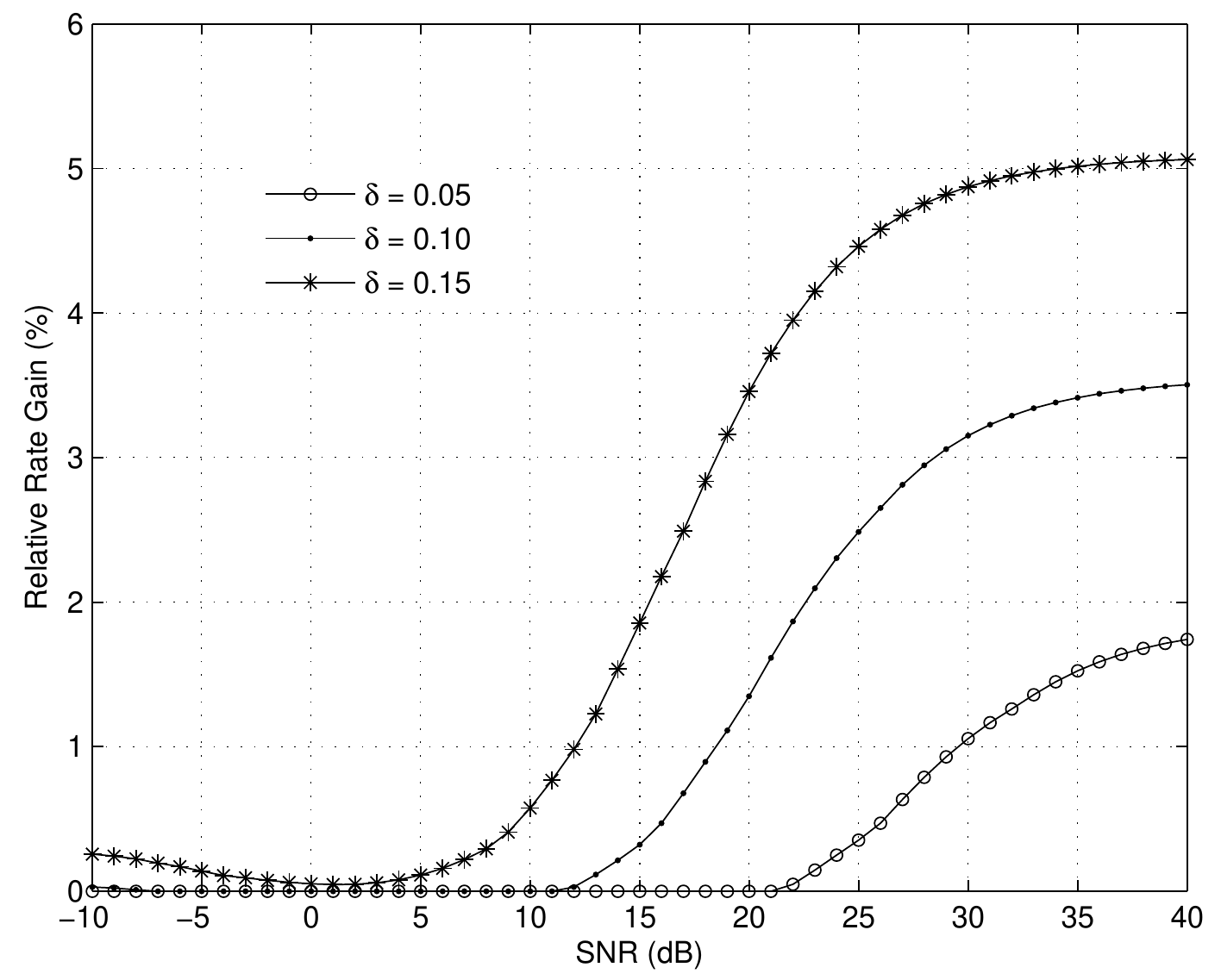} \vskip-2mm
\caption{The relative rate gain by adopting the optimal training sequence length, compared to the case of using training sequence of length $N_t$ ($N_t = N_r = 4, T = 100$).} \label{fig:performancegain} \vskip-3mm
\end{centering}
\end{figure}

In Fig. \ref{fig:CAP_ALFA}, we have plotted the approximated achievable rate with the optimal power allocation scheme. For each SNR value, we choose the best training sequence length $T_p^{\mathrm{opt}}$. It is noteworthy that, for the hardware impaired systems, the achievable rate saturates when SNR becomes high, even though we have used the optimized scheme. This behavior remains even if we have perfect CSI as in \cite{Emil2013imp}, thus it is a fundamental effect of hardware impairments. In Fig. \ref{fig:performancegain}, we plot the $\textit{relative~rate~gain}$ by adopting the optimal training sequence length $T_p^{\mathrm{opt}}$. The relative rate gain is defined as
\begin{equation}
\textrm{relative~rate~gain} \triangleq \frac{R_{T_p^{\mathrm{opt}}} - R_{T_p = N_t}}{R_{T_p= N_t}}\times 100\%,
\end{equation}
where $R_{T_p^{\mathrm{opt}}}$ and $R_{T_p = N_t}$ refer to the approximated achievable rate (\ref{eq:analcapacity}) when $T_p$ obtains its optimal value and $T_p\!=\!N_t$, respectively.
We can conclude from this figure that, the relative rate gain provided by utilizing the optimal training sequence length, varies according to the level of RTRI. Systems with higher level of impairments benefit far more from the optimization over $T_p$.


\subsection{Equal Training and Data Power} \label{sec:equalpower}

In practice, communication systems do not often have the freedom of varying the transmit powers during the training phase and data transmission phase. As such, the transmit power for pilot and data symbols is always the same, i.e., $\rho_p = \rho_d = \rho$. In this case, the effective SNR in (\ref{eq:effSNR}) becomes
\begin{equation}
\rho_{\mathrm{eff}} = \frac{\rho^2T_p}{N_t(1+\rho\delta^2)(1+\rho+\rho\delta^2)+(\rho^2\delta^2+\rho)T_p}.\label{eq:effSNR_NOALFA}
\end{equation}
The corresponding analytical approximated achievable rate follows straightforward by inserting (\ref{eq:effSNR_NOALFA}) into (\ref{eq:analcapacity}). Using the obtained analytical rate expression, we can, once more, resort to exhaustive search to find the optimal training sequence length.

\begin{figure}[Ht!]
\begin{centering}
\includegraphics [keepaspectratio,width=\columnwidth]{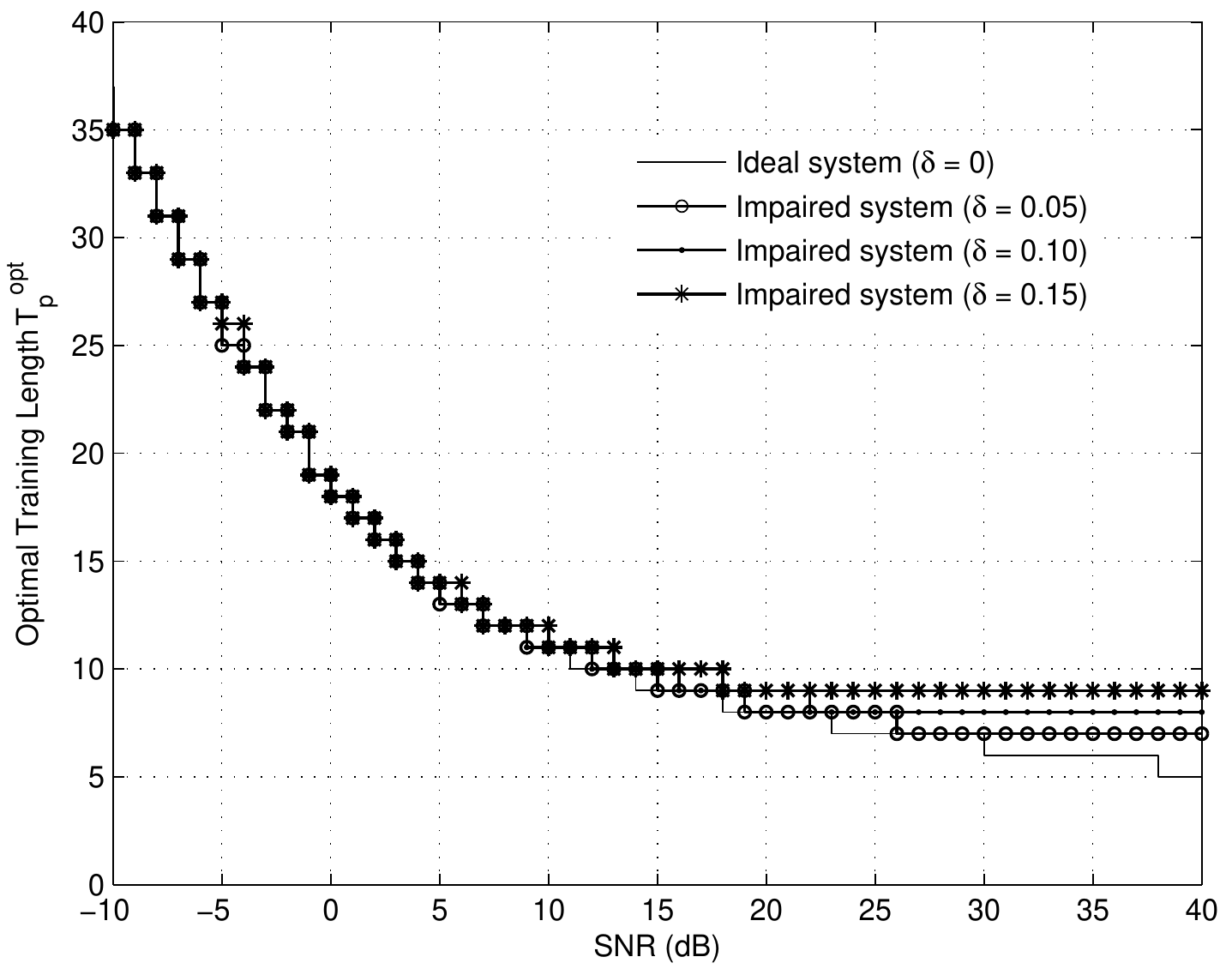} \vskip-2mm
\caption{Optimal training sequence length for different levels of impairments when only the training sequence length is optimized ($N_t = N_r = 4, T = 100, \rho_p = \rho_d = \rho$).} \label{fig:Tp_NOALFA} \vskip-3mm
\end{centering}
\end{figure}

Figure \ref{fig:Tp_NOALFA} depicts the optimal $T_p$ for a $4\times 4$ MIMO system with coherence time $T = 100$. As we can see, for all cases, the demand for training is especially high at low SNRs, whilst this demand decreases as the SNR scales up.
Generally speaking, higher level of RTRI require longer training length in the high SNR regime, whereas such demands are nearly the same for all practical levels of impairments at low SNRs.

\section{Conclusions}
In this paper, we analyzed the impact of residual transmit RF impairments on training-based MIMO systems. We derived a new LMMSE channel estimator for systems with RTRI, and then found that such residual impairments create an irreducible estimation error floor. Moreover, the optimal power allocation scheme and optimal training sequence length were thereafter investigated. We showed that the optimal training sequence length may be larger than the number of transmit antennas, and increases with the level of impairments. An increase in the relative rate is observed by adopting the optimal training sequence length. We also investigated the optimal training sequence length when there is no freedom of varying the transmit power during the estimation and data transmission phases, and concluded that the demand for training is the same at low SNRs, while more training was needed at high SNRs when the system experiences RTRI.

\section*{ACKNOWLEDGMENTS}
The work of X.~Zhang, M.~Matthaiou and M. Coldrey has been supported in part by the Swedish Governmental Agency for Innovation Systems (VINNOVA) within the VINN Excellence Center Chase, and by the Swedish Foundation for Strategic Research. The work of E.~Bj{\"o}rnson has been supported by the International Postdoc Grant 2012-228 from the Swedish Research Council, and by the ERC Starting Grant 305123 MORE.

\bibliographystyle{IEEEtran}
\bibliography{IEEEabrv,Reference}

\begin{thebibliography}{10}
\providecommand{\url}[1]{#1}
\csname url@samestyle\endcsname
\providecommand{\newblock}{\relax}
\providecommand{\bibinfo}[2]{#2}
\providecommand{\BIBentrySTDinterwordspacing}{\spaceskip=0pt\relax}
\providecommand{\BIBentryALTinterwordstretchfactor}{4}
\providecommand{\BIBentryALTinterwordspacing}{\spaceskip=\fontdimen2\font plus
\BIBentryALTinterwordstretchfactor\fontdimen3\font minus
  \fontdimen4\font\relax}
\providecommand{\BIBforeignlanguage}[2]{{%
\expandafter\ifx\csname l@#1\endcsname\relax
\typeout{** WARNING: IEEEtran.bst: No hyphenation pattern has been}%
\typeout{** loaded for the language `#1'. Using the pattern for}%
\typeout{** the default language instead.}%
\else
\language=\csname l@#1\endcsname
\fi
#2}}
\providecommand{\BIBdecl}{\relax}
\BIBdecl

\bibitem{telatar1999capacity}
E.~Telatar, ``Capacity of multi-antenna {Gaussian} channels,'' \emph{Europ.
  Trans. Telecom.}, vol.~10, no.~6, pp. 585--595, Nov.-Dec. 1999.

\bibitem{foschini1998a}
G.~J. Foschini and M.~J. Gans, ``On limits of wireless communications in a
  fading environment when using multiple antennas,'' \emph{Wireless Pers.
  Commun.}, vol.~6, no.~3, pp. 311--335, Mar. 1998.

\bibitem{hassibi2003}
B.~Hassibi and B.~M. Hochwald, ``How much training is needed in
  multiple-antenna wireless links?'' \emph{{IEEE} Trans. Inf. Theory}, vol.~49,
  no.~4, pp. 951--963, Apr. 2003.

\bibitem{tong2004pilot}
L.~Tong, B.~M. Sadler, and M.~Dong, ``Pilot-assisted wireless transmissions:
  {G}eneral model, design criteria, and signal processing,'' \emph{{IEEE}
  Signal Process. Mag.}, vol.~21, no.~6, pp. 12--25, Nov. 2004.

\bibitem{biguesh2006training}
M.~Biguesh and A.~B. Gershman, ``Training-based {MIMO} channel estimation: {A}
  study of estimator tradeoffs and optimal training signals,'' \emph{{IEEE}
  Trans. Signal Process.}, vol.~54, no.~3, pp. 884--893, Mar. 2006.

\bibitem{mikael1}
M.~Coldrey and P.~Bohlin, ``Training-based {MIMO} systems--{P}art {I}:
  {P}erformance comparison,'' \emph{{IEEE} Trans. Signal Process.}, vol.~55,
  no.~11, pp. 5464--5476, Nov. 2007.

\bibitem{emil2010est}
E.~Bj{\"{o}}rnson and B.~Ottersten, ``A framework for training-based estimation
  in arbitrarily correlated {R}ician {MIMO} channels with {R}ician
  disturbance,'' \emph{{IEEE} Trans. Signal Process.}, vol.~58, no.~3, pp.
  1807--1820, Nov. 2010.

\bibitem{agarwal2012adaptive}
M.~Agarwal, M.~L. Honig, and B.~Ata, ``Adaptive training for correlated fading
  channels with feedback,'' \emph{{IEEE} Trans. Inf. Theory}, vol.~58, no.~8,
  pp. 5398--5417, Aug. 2012.

\bibitem{schenk2008rf}
T.~Schenk, \emph{RF Imperfections in High-Rate Wireless Systems: Impact and
  Digital Compensation}.\hskip 1em plus 0.5em minus 0.4em\relax Springer, 2008.

\bibitem{studer2010residual}
C.~Studer, M.~Wenk, and A.~Burg, ``{MIMO} transmission with residual
  transmit-{RF} impairments,'' in \emph{Proc.~ITG/IEEE Work. Smart Ant. (WSA)},
  Feb. 2010, pp. 189--196.

\bibitem{emil2013est}
E.~Bj{\"{o}}rnson, J.~Hoydis, M.~Kountouris, and M.~Debbah, ``Massive {MIMO}
  systems with non-ideal hardware: Energy efficiency, estimation, and capacity
  limits,'' \emph{{IEEE} Trans. Inf. Theory}, 2013, submitted, arXiv:1307.2584.

\bibitem{holma2011LTE}
H.~Holma and A.~Toskala, \emph{{LTE} for {UMTS}: {E}volution to
  {LTE}-{A}dvanced}.\hskip 1em plus 0.5em minus 0.4em\relax Wiley, 2011.

\bibitem{don2012double}
N.~O'Donoughue and J.~Moura, ``On the product of independent complex
  {G}aussians,'' \emph{{IEEE} Trans. Signal Process.}, vol.~60, no.~3, pp.
  1050--1063, Mar. 2012.

\bibitem{bertsekas1999nonlinear}
D.~P. Bertsekas, \emph{Nonlinear programming}.\hskip 1em plus 0.5em minus
  0.4em\relax Athena Scientific, 1999.

\bibitem{kaybook1}
S.~M. Kay, \emph{Fundamentals of Statistical Signal Processing, Volume 1:
  Estimation theory}.\hskip 1em plus 0.5em minus 0.4em\relax Prentice Hall PTR,
  1993.

\bibitem{gradshteyn2007a}
I.~S. Gradshteyn and I.~M. Ryzhik, \emph{Table of Integrals, Series, and
  Products}, 7th~ed.\hskip 1em plus 0.5em minus 0.4em\relax Academic Press,
  2007.

\bibitem{zanella2009marginal}
A.~Zanella, M.~Chiani, and M.~Z. Win, ``On the marginal distribution of the
  eigenvalues of {Wishart} matrices,'' \emph{{IEEE} Trans. Commun.}, vol.~57,
  no.~4, pp. 1050--1060, Apr. 2009.

\bibitem{kang2006rician}
M.~Kang and M.-S. Alouini, ``Capacity of {MIMO} {R}ician channels,''
  \emph{{IEEE} Trans. Wireless Commun.}, vol.~5, no.~1, pp. 112--122, Jan.
  2006.

\bibitem{Emil2013imp}
E.~Bj{\"{o}}rnson, P.~Zetterberg, M.~Bengtsson, and B.~Ottersten, ``Capacity
  limits and multiplexing gains of {MIMO} channels with transceiver
  impairments,'' \emph{{IEEE} Commun. Lett.}, vol.~17, no.~1, pp. 91--94, Jan.
  2013.

\end{thebibliography}

\end{document}